\documentclass[aps,longbibliography,notitlepage,nofootinbib,11pt,floatfix,tightenlines
]{revtex4-2}

\usepackage{bm}
\usepackage{amsmath}
\usepackage{graphicx}
\usepackage{graphics}
\usepackage{xcolor}
\usepackage{mathtools}
\usepackage{amsfonts}
\usepackage{array}
\usepackage{empheq}
\usepackage{multirow}
\usepackage{tabularx}

\usepackage[format=plain,labelfont={bf,small},textfont=small,justification=raggedright,singlelinecheck=false]{caption}
\newcommand{\sS}{{\scriptscriptstyle S}}
\newcommand{\sN}{{\scriptscriptstyle N}}
\newcommand{\sT}{{\scriptscriptstyle T}}

\begin{document}

\author{J. A. Hanna}\email{jhanna@unr.edu}
\author{R. S. Hutton}
\affiliation{Mechanical Engineering, University of Nevada,
     1664  N.  Virginia  St.\ (0312),  Reno,  NV  89557-0312,  U.S.A.}
 
 \title{A derivation of viscous thin film flow equations on curved surfaces}

\begin{abstract}
General equations are derived for slow viscous thin fluid film flows on curved surfaces through an extension of Leal's pedagogical approach, which leaves the characteristic velocity scale unspecified and employs a direct through-thickness integration of the continuity equation. The derivation neglects inertia, and includes gravitational, capillary, and Marangoni effects, the latter coupling the thickness dynamics to free-surface transport of a dilute, non-diffusing surfactant. The resulting general expression incorporates the leading order terms of each type, as well as additional terms that become leading order for nongeneric cases.  A few examples are briefly presented and literature comparisons made.  The importance of gradients in curvature is emphasized, and it is suggested that nondimensionalization of geometric features might lead to further useful generalizations. This relatively simple formulation is intended as a starting point for exploring interactions between geometry, gravity, and surface tension. 
\end{abstract}

\date{\today}

\maketitle

\section{Introduction}\label{sec:introduction}

Thin film flows of a viscous fluid with one free surface are encountered across a range of scales in geological, industrial, culinary, biological, and other settings. General reviews, mostly in a flat geometric context, can be found in \cite{Myers98, Oron97, CrasterMatar09}. 
The intent of the current work is to facilitate investigation of the role of substrate geometry in driving, maintaining, or modulating such flows. 
In that light, we do not review the extensive literature on flows over planar surfaces, nor vertical flows over fibers or inside tubes, surfaces that are uniform in the direction of motion. We also restrict our discussion to surfaces without a secondary small dimension, which excludes things such as narrow grooves. There are many derivations of governing equations for the thickness of a film on curved surfaces, examples of which follow, incorporating one or more effects such as inertia \cite{Pukhnachev77, PreziosiJoseph88, Roy02, Weidner18, SahuKumar14, LiKumar18, Parrish22, Wray17, Duruk23, Shepherd24}, gravity \cite{Moffatt77, Pukhnachev77, PreziosiJoseph88, ReisfeldBankoff92, Weidner13, Weidner18, RoySchwartz97, Roy02, Myers02, thiffeaultkamhawi06, TakagiHuppert10, SahuKumar14, LiKumar18, Parrish22, Balestra16, Kang17, Wray17, Shi20, XueStone21, Duruk23, Shepherd24}, capillarity (surface tension) \cite{Pukhnachev77, PreziosiJoseph88, ReisfeldBankoff92, SchwartzWeidner95, Weidner13, Weidner18, Myers02, RoySchwartz97, Roy02, Howell03, thiffeaultkamhawi06, SahuKumar14, LiKumar18, Parrish22, Kang17, Wray17, Shi20, Duruk23, Shepherd24}, and Marangoni (gradient of surface tension) forces, the last being coupled to either position- or height-dependent \cite{ReisfeldBankoff92, Kang17} 
or dynamic \cite{Weidner13, LiKumar18, Shi20} fields of active scalars such as temperature or surfactant concentration. Other complexities include van der Waals forces \cite{ReisfeldBankoff92}, a dynamic substrate \cite{Howell03}, solidification \cite{Myers02}, and condensation or evaporation \cite{Myers02, Shi20}.  Geometry has also been examined in a small-slope context, for example in \cite{HwangMa89, StillwagonLarson90, Pritchard92, DuffyMoffatt95, Kalliadasis00, Gramlich02, Gaskell04, Hinton19}, and for steady flows by qualitatively different approaches employing Green's functions \cite{Hayes00, BlythPozrikidis06}. 

We are concerned with a regime in which the geometry is not approximated. 
Many studies have considered simple curved shapes such as circular cylinders or spheres which, while common, are exceptional cases from the point of view of capillarity-driven flows, as they do not feature gradients in curvature. 
The dominant capillary term comes from the gradient of mean curvature, so the scaling of geometric terms for such shapes is nongeneric, thus potentially misleading if used to exemplify and study curvature effects, or confusing if one compares equations across different specific examples of surfaces.  A few studies consider more complex shapes such as ellipsoids or tori, general one-dimensional, cylindrical, or axisymmetric surfaces, or cones.   A few exceptional works derive equations for general surface geometries \cite{Howell03, Myers02, RoySchwartz97, Roy02, thiffeaultkamhawi06, Duruk23, Shepherd24}.  Howell \cite{Howell03} uses lines of curvature coordinates, and includes only capillary forces. He provides a thoughtful analysis of different geometric regimes, noting for example that constant curvature is its own regime. The resulting equations are simple enough to be analytically usable, particularly after some higher order terms are discarded. Myers, Charpin, and Chapman \cite{Myers02} also use lines of curvature coordinates, and include gravity and capillarity along with other effects relevant to accretion. They also briefly but thoughtfully address different geometric regimes of capillarity-driven flow.  These equations are also usable, particularly when accretion is not included.  
Some other derivations result in very complex equations. Roy and Schwartz \cite{RoySchwartz97} use lines of curvature coordinates, and include gravity and capillarity.  This work is less often cited than that of Roy, Roberts, and Simpson \cite{Roy02} which builds upon it, using general coordinates and adding inertial effects, but which is somewhat opaque in its derivation, invoking the use of a computer algebra program. 
Thiffeault and Kamhawi \cite{thiffeaultkamhawi06} use general coordinates, and include both gravity and capillarity. Their work shares some aspects of our kinematic approach, but being differently motivated, they did not attempt to reduce the equations into their simplest form. 
Recent works by Duruk, Shepherd, Boujo, and Sellier \cite{Duruk23, Shepherd24} built on these derivations, with the important addition of inertial terms from rotation of the substrate around an arbitrary axis.  
The result of such derivations is bewildering yet non-exhaustive collections of perturbative terms of different orders scattered throughout the equations, whose physical interpretation and interaction with other effects is difficult to understand. It is not obvious 
 which terms might be neglectable, and indeed several others have simply adopted such equations whole. 
For example, the equations of \cite{Roy02}, with or without simplifications suggested by \cite{Howell03}, have been applied to explore particular classes of shapes in \cite{Braun12, Dukler20, Lin21, Ledda22}. 
 Given the complexity of such equations, it is typical to resort either to a numerical solution, or a severe analytical approximation of the geometry akin to a small-slope assumption. This situation makes it difficult to address the question of the effects of geometry, including its interaction with other physics.

It is our intent to suggest that a less unhappy medium is possible, one that should allow for some analytical probing of the effects of substrate geometry in general coordinates. Additionally, we are unaware of any treatment of thin film equations featuring coupling between an evolving Marangoni field and a general substrate, although others have considered cylinders or spherical caps (for example \cite{Weidner13, LiKumar18, Shi20}) or fixed thermally-driven leveling over gentle topography \cite{Gramlich02}, and there are relevant examples in other settings, such as the stability of films with two free surfaces \cite{IdaMiksis98-1,*IdaMiksis98-2} or the breakup of drops and bubbles in flows \cite{Stone94review}. 

The outcome of the present work connects with a variety of prior results, but the approach differs in several subtle ways. 
We follow and extend the pedagogical approach of Leal's textbook \cite[Sections 5B and 6A]{leal2007advanced}, which we adapt to a curved substrate. There are two particularly notable aspects of this approach. One is that there are no initial assumptions on the characteristic velocity scale, which is left to be specified later when driving forces are selected for modeling. This allows for a single general equation incorporating several potentially relevant forces. The other is that it employs a simple through-thickness integration of the continuity equation, which we find is still an appropriate approximation at leading geometric order.  
This avoids calculations that embed additional higher order terms throughout the equations, including nonlinearities in the time derivatives. 
We wish to capture the leading terms in both generic and special cases, as the latter are quite commonly encountered for particularly simple geometries. This should enable us to study and understand the role of substrate geometry and how it interacts with other effects such as gravitational, capillary, and Marangoni forces, in a general setting rather than requiring \emph{ad hoc} equations for each orientation and geometry.  Missing from our derivation, however, is the role of inertia, which is often seen in examples such as the rotation of a cylinder, and   is given a more general geometric treatment in \cite{Duruk23, Shepherd24}. 

This paper is organized as follows.  
In Section \ref{sec:setup}, we introduce our notation and coordinate system, the lubrication equations and boundary conditions, and an expansion in geometric quantities distinct from the classical asymptotic expansion of the fields in the lubrication approximation. 
In Section \ref{sec:evolution}, we derive an evolution equation for the thickness of the thin film, retaining only the leading order terms of each type, with allowance for special cases in which the relative orders of terms follow nongeneric scalings. 
In Section \ref{sec:surfactant}, we introduce the simplest possible model of a dynamic surfactant field on the free surface. This allows us to finalize, in Section \ref{sec:groups}, the physical dimensionless groups, and present general coupled evolution Equations (\ref{dhdtnon}-\ref{usnh}) for the thickness and surfactant concentration, which are displayed in their most usably simple generic form as Equations (\ref{dhdtgeneric}-\ref{surftransgeneric}) in Section \ref{sec:generic}.  These two sets of equations are the primary result of the paper.  In Section \ref{sec:examples}, we provide examples of the form of the equations in a few different regimes, and briefly compare with some of the relevant literature. Discussion follows in Section \ref{sec:discussion}.

\section{Formulation} \label{sec:setup}

\subsection{Preliminaries}\label{sec:preliminaries}

This subsection will be the only part of the paper where quantities, both physical and geometric, carry dimensions.  
To avoid using diacritics or scripts, the same symbols will be re-used when these quantities appear as their dimensionless selves in the rest of the paper. 

The incompressible Navier-Stokes equations for a fluid are written in Eulerian form, 
\begin{align}
\rho \left( \frac{\partial \bm{u}}{\partial t} + \bm{u} \cdot {\nabla}\bm{u} \right) &= \rho \bm{G} -{\nabla}p +\mu {\nabla}^2 \bm{u} \label{NS} \, , \\
{\nabla} \cdot \bm{u} &= 0 \, , \label{continuity}
\end{align}
where $\bm{u}$ and $p$ are velocity and pressure fields, $t$ is time, $\bm{G}$ is the gravitational acceleration, $\rho$ and $\mu$ are 
density and dynamic viscosity. These equations will be reduced by assuming a thin viscous film, and adapting the standard lubrication assumptions to a curved geometry.   This aspect of our approach is similar to that of \cite{thiffeaultkamhawi06}. 

We construct a fixed, surface-adapted coordinate system that conforms to a fixed solid surface, the substrate. There are two tangential coordinates $\eta^\alpha$, $\alpha \in \{1,2\}$, with corresponding derivatives $\partial_\alpha$ and coordinate basis vectors $\bm{g}_\alpha$. Orthogonal to these is a normal through-thickness coordinate $N \ge 0$ and an outward-pointing \emph{unit} normal $\bm{N}$.  The tangent basis is a function of all three coordinates, while the unit normal is only a function of the tangential coordinates. Reciprocal tangent basis vectors are defined such that $\bm{g}^\alpha\cdot\bm{g}_\beta = \delta^\alpha_\beta$. Indices are raised and lowered with the metric and inverse metric components, defined in the usual way with these bases, along with the gradient $\nabla$ and the covariant derivative.  On the substrate, $N=0$ and \(\bm{g}_\alpha |_{N=0} = \bm{a}_\alpha\), the substrate tangent basis, which can accordingly be used to define a reciprocal basis $\bm{a}^\alpha$ and associated quantities such as the metric and covariant derivative $\nabla_\alpha$ on the substrate.  
To avoid potential confusion, we will only write the component form of substrate quantities, rather than quantities elevated off of the surface (what \cite{thiffeaultkamhawi06} call in the ``bulk''). We will need the substrate curvature tensor $\bm{b} = b_{\alpha\beta}\bm{a}^\alpha\bm{a}^\beta$, its components $b_{\alpha\beta} = b_{\beta\alpha} = \partial_\alpha\bm{a}_\beta\cdot\bm{N}=-\bm{a}_\beta\cdot \partial_\alpha\bm{N}$, and its invariants the mean $H$ and Gau{\ss}ian $K$ curvatures, where $2H=b^\alpha_\alpha$ and $2K=b^\alpha_\alpha b^\beta_\beta - b^\alpha_\beta b_\alpha^\beta$, 
as well as the Weingarten relations $\nabla_\beta\bm{a}_\alpha = b_{\alpha\beta}\bm{N}$ and $\partial_\alpha\bm{N} = -b_{\alpha\beta}\bm{a}^\beta$, and the Codazzi relations $\nabla_\gamma b_{\alpha\beta} = \nabla_\alpha b_{\gamma\beta}$. 
Note that while individual tangent or reciprocal tangent basis vectors depend on the through-thickness coordinate $N$, decomposition into tangential and normal parts can be effected with any parallel tangent plane, 
 independently of $N$, as the identity tensor can be written either as $\bm{g}_\alpha\bm{g}^\alpha + \bm{N}\bm{N}$ or $\bm{a}_\alpha\bm{a}^\alpha + \bm{N}\bm{N}$. However, the gradient operator $\nabla$ 
  depends on $N$. 

The system will be thin in the normal direction.  Elevated operators and geometric quantities will be expressed using the substrate basis through standard geometric expansion techniques \cite{Flugge72}. This is distinct from the classical asymptotic expansion of physical quantities based on lubrication assumptions, although both are consequences of the thin geometry. 

\begin{figure}[h]
    \centering
    \includegraphics[width=0.5\textwidth]{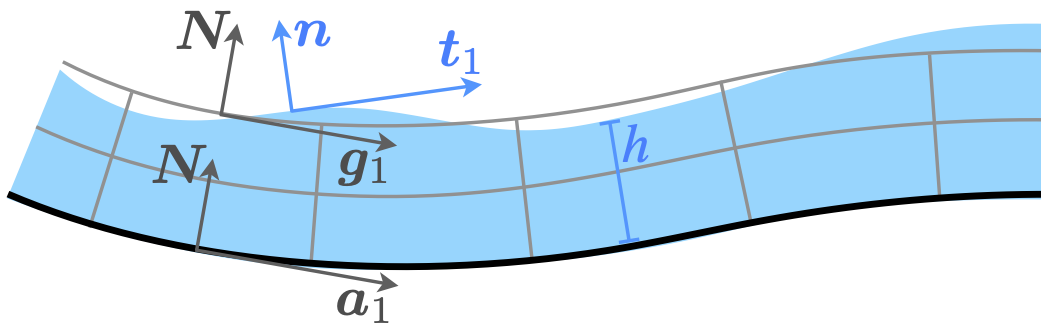}
    \caption{The coordinate system. Fixed substrate (black curve) with tangents $\bm{a}_\alpha$ and unit normal $\bm{N}$, fixed adapted coordinates (grey grid) with tangents $\bm{g}_\alpha$ and the same normals, and fluid (light blue region) with local thickness $h$ and free surface tangents $\bm{t}_\alpha$ and unit normal $\bm{n}$. Only one tangential direction is shown for simplicity, but quantities may vary, and vectors may rotate, into the page.}
    \label{coordinatesystem}
\end{figure}

Moving through this coordinate system is a fluid with one boundary on the substrate and one boundary free, at height $N=h$ with unit normal $\bm{n}$, where $h$ and $\bm{n}$ are functions only of the tangential coordinates, and can be safely assumed unique given more restrictive conditions implied by the lubrication approximation. 
These are schematically shown in Figure \ref{coordinatesystem}, along with a free boundary tangent $\bm{t}_\alpha$, which will be effectively defined in the next subsection.  

\subsection{Scalings, expansions, and more notation}\label{sec:scalings}

The thin geometry results in a separation of scales tangent and normal to the substrate, which we now use to rescale physical and geometric quantities. All equations from here onward are nondimensional unless otherwise specified. 

The ratio of a characteristic thickness $h_c$ to a characteristic tangential length scale $l_c$ is a small parameter $\varepsilon \equiv h_c/l_c$.  Tangential distances and gradient will be scaled and inverse scaled by $l_c$, while thickness $h$, through-thickness distance $N$ and its derivative will use $h_c = \varepsilon l_c$.  However, a curved geometry introduces at least one other length scale, a characteristic radius of curvature $R_c$. This embodies another way by which quantities change in the tangential direction, through the geometry, distinct from how they might change in the absence of substrate curvature. This additional length will scale the curvatures.  Note that it is the full curvature tensor or mixed components of curvature, and thus the invariants, that are scaled with $R_c$, just as it is the full tangential gradient, and not its bare components (derivatives) or basis vectors, that are scaled with $l_c$. In addition to the small parameter $\varepsilon = h_c/l_c$, there is thus another small parameter $h_c/R_c$, which will be written throughout as $\varepsilon l_c/R_c$.  To maintain the legitimacy of the lubrication approximation, the geometric scale $R_c$ should be of an order no smaller than that of the tangential scale $l_c$. While $l_c/R_c$ can be no larger than $O(1)$, it can be smaller, and vanishes for planar surfaces.  Using a separate length scale $R_c$ must be reconciled with the description of the geometry of the substrate, which additional step will arise when considering particular geometries. Our present discussion provides only a minimum of detail, and we assume it will always be possible to do this. For example, a position vector or height function of the surface might be broken into a flat part and a variable part that is scaled with a characteristic radius. For a cylinder or channel, it might be sensible to further rescale the tangential gradient differently in its two directions. We will also see later that capillary physics suggests further consideration of the scaling of mean curvature and its variation, which might warrant the introduction of a new geometric nondimensional number, analogous to the manner in which capillary and Marangoni forces are treated separately.  

Tangential velocities are nondimensionalized by a characteristic scale $u_c$ which, following Leal \cite{leal2007advanced}, we leave unspecified to allow different possible force balances to be selected later. It is implied that this scale will be small, carrying some powers of $\varepsilon$.  The normal velocity is one order smaller, and is scaled by $\varepsilon u_c$. Derived from the above are a characteristic time $t_c = l_c/u_c$ and pressure $p_c =  \mu u_c / {\varepsilon}^2 l_c$, anticipating the standard lubrication balances.  Some other scalings are obvious, including gravity by the standard magnitude of gravitational acceleration $G$, and surface tension, which will appear in Section \ref{sec:eqsandbc}.  Surfactant concentration will appear in Section \ref{sec:surfactant}. 

Operators and representations of fields are split because of the different distance and velocity rescalings in tangential and normal directions. 
For example, 
$\bm{u} = \bm{u}_\sT + \varepsilon u_\sN \bm{N}$ and ${\nabla} = \nabla_\sT + \tfrac{1}{\varepsilon}\bm{N}\partial_\sN$.  
In contrast, this does not happen to something without such scales, like gravity $\bm{G} = \bm{G}_\sT  + G_\sN \bm{N}$. 
It should be clear from context throughout that Latin subscripts are distinct from Greek indices, and only the latter follow the summation convention. 

There are now two expansions to consider.  One follows the standard lubrication approach, in which the velocity and pressure fields are given a regular asymptotic expansion in $\varepsilon$. This is kept implicit in the present paper, in which only the lowest order terms in velocity and pressure will be retained.  To avoid cumbersome notation, as well as to treat both expansions in a similar way, we indicate the truncation of this standard expansion only by using $O$ notation to append the orders of dropped terms in each expression, rather than using a scripted zero or other label on the velocities and pressures.   The other is a Taylor expansion in $\varepsilon N$ of the geometry of the coordinate system in the vicinity of the substrate. This will always involve the rescaled curvatures, so will always appear as an expansion in $\varepsilon\tfrac{l_c}{R_c}$.  We will keep separate track of this geometric small parameter, so one can tell what approximations are coming from the asymptotic assumption or from the proper Taylor expansion.  There will result terms that have products of these two small parameters. As the asymptotic expansion is on a different footing than the geometric one, this is an informal abuse of $O$ notation, but one that can be unraveled if desired.   We believe something similar is being done in \cite{Howell03}. 
 
The geometric expansions \cite{Flugge72} incorporating the characteristic radius involve expanding the tangents around the substrate. From the rescaled but exact relation for the tangents\newline $\bm{g}_\alpha = \bm{a}_\alpha - \varepsilon \tfrac{l_c}{R_c} N b_\alpha^\beta \bm{a}_\beta$, one obtains the truncated expansion of the inverse tangents\newline $\bm{g}^\alpha = \bm{a}^\alpha + ( \varepsilon \tfrac{l_c}{R_c} N b^\alpha_\beta + O(\varepsilon^2 \tfrac{l_c^2}{R_c^2}) ) \bm{a}^\beta$.  The latter expression uses a shorthand indicating merely that the truncated part is still tangential, not that it is proportional to the tangent, as the dropped piece in parentheses also carries indices and can rotate the tangent in the tangent plane in addition to changing its length. The gradient on the substrate will be written $\nabla_\sS \equiv \nabla_\sT |_{N=0} = \bm{a}^\alpha \partial_\alpha$. This substrate gradient will be used to approximate the tangential part of the gradient acting on elevated quantities. This will facilitate a through-thickness integration to express everything on the substrate surface, as it leaves only the physical fields $\bm{u}_\sT$, $u_\sN$, and $p$ dependent on $N$.  The pieces $\bm{G}_\sT$,  $G_\sN$, and $\bm{N}$ appearing in gravity are also independent of $N$, and can be treated like surface quantities. As it is understood that the physical fields are the leading order terms in an asymptotic expansion (in more typical notation, these symbols would carry a zero), our expressions will represent the leading order terms in \emph{both} expansions, approximating both physical and geometric quantities. With this notation we can, for example, write the divergence of the velocity field in the form $ \nabla \cdot \bm{u} = \nabla_\sS \cdot \bm{u}_\sT + \partial_\sN u_\sN + O(\varepsilon, \varepsilon\tfrac{l_c}{R_c})$, 
where the $O$ terms include higher order terms in the asymptotic expansion of the velocity, a curvature term coupling to the normal velocity, and terms associated with the use of the substrate gradient.  Writing the dropped asymptotic terms as $O(\varepsilon)$ may be overly conservative, given that the classical lubrication approximation would indicate $O(\varepsilon^2)$ (and see also the treatment in \cite{Howell03}), but as noted in \cite{Myers02}, this simplifies the development. While expressions such as $\nabla_\sS \cdot \bm{u}_\sT$ or $\bm{u}_\sT \cdot \nabla_\sS$ appearing below might seem at first glance ambiguous to implement, recall that the velocity (which is itself dependent on $N$), can be resolved equally onto $\bm{g}_\alpha\bm{g}^\alpha$ or $\bm{a}_\alpha\bm{a}^\alpha$, which are independent of $N$. 

While no approximations are made to the substrate geometry, the lubrication assumptions motivating the rescaling of $h$ imply small deviations of the slope of the free boundary from that of the substrate. This is more restrictive than the requirement that $h$ and other free surface quantities be unique. Let a function $F \equiv \varepsilon(N-h)$ vanish on the free surface, hence its gradient $\nabla F = \bm{N} - \varepsilon\nabla_\sT h = \bm{N} -  (\varepsilon + O(\varepsilon^2 \tfrac{l_c}{R_c})) \nabla_\sS h$ will be normal to the free surface, again using a shorthand as the dropped term expresses the action of the curvature which can both rotate and stretch $\nabla_\sS h$ in the tangent plane. The free boundary tangents $\bm{t}_\alpha$ and unit normal $\bm{n}$ can either be determined from the free surface position (located $h\bm{N}$ away from the corresponding substrate position) or expanded around corresponding fixed basis vectors $\bm{g}_\alpha$ and $\bm{N}$, 
\begin{align}
	\bm{n} = \frac{\nabla F}{| \nabla F |} &= (1+O(\varepsilon^2)) ( \bm{N} - \varepsilon\nabla_\sT h |_{N=h} ) 
	= (1+O(\varepsilon^2))\bm{N} - (\varepsilon + O(\varepsilon^2 \tfrac{l_c}{R_c},\varepsilon^3) )\nabla_\sS h \, , \label{surfnormal} \\
	\bm{t}_\alpha &= \bm{g}_\alpha |_{N=h} + \varepsilon \partial_\alpha h \bm{N} 
	= \bm{a}_\alpha - \varepsilon\tfrac{l_c}{R_c}hb_\alpha^\beta\bm{a}_\beta + \varepsilon \partial_\alpha h \bm{N} \, ,	\\
	\bm{t}^\alpha &= \bm{a}^\alpha + ( \varepsilon \tfrac{l_c}{R_c} h b^\alpha_\beta + O(\varepsilon^2 \tfrac{l_c^2}{R_c^2}, \varepsilon^2) ) \bm{a}^\beta  + (\varepsilon + O(\varepsilon^3) ) \nabla^\alpha h \bm{N} \, ,\label{surftangent}
\end{align}
(shorthand applies), where $\nabla^\alpha$ uses the substrate metric. 
Our treatment of orders does not distinguish between $h$ and its gradient.  
An important quantity appearing in capillary terms is the trace of free surface curvature
\begin{align}
-\nabla \cdot \bm{n} &=  -\bm{t}^\alpha\cdot \partial_\alpha \bm{n} = -(\bm{a}^\alpha + \varepsilon\tfrac{l_c}{R_c}hb^\alpha_\beta\bm{a}^\beta +\varepsilon\nabla^\alpha h\bm{N}) \cdot \partial_\alpha (\bm{N} - \varepsilon \nabla_\sS h ) + O(\varepsilon^2 \tfrac{l_c}{R_c},\varepsilon^3) \nonumber \\
&= \tfrac{l_c}{R_c}2H + \varepsilon \tfrac{l_c^2}{R_c^2}\, h (4H^2 - 2K)  + {\varepsilon}\nabla^2_\sS h + O(\varepsilon^2 \tfrac{l_c}{R_c},\varepsilon^3) \, , \label{surfdivn}
\end{align}
where \mbox{$\nabla^2_\sS \equiv \nabla_\sS \cdot \nabla_\sS$}, and Weingarten has been used in its rescaled form 
\mbox{$\partial_\alpha \bm{N} = -\tfrac{l_c}{R_c}b_\alpha^\beta\bm{a}_\beta$}. 
The first new term resulting from substrate curvature is at order one.  The gradient of this term will appear in the final equations, as these terms will find their way into the pressure through a boundary condition, and the gradient of the pressure will contribute to the velocity. This is the largest and thus most important term, unless the substrate is a surface of constant mean curvature, as many prior examples have been---  planes, cylinders, and spheres. Because of this possibility, two terms are retained at next order, one a coupling term coming from the geometric expansion of the mean curvature of the coordinate lines off of the substrate, and one from distortions of the free surface that is also present in the standard lubrication equations for flat substrates.  The truncation for curved substrates is at a lower order than for flat substrates.

\subsection{Lubrication equations and boundary conditions}\label{sec:eqsandbc}

We are now ready to proceed analogously to the standard way, from the momentum and continuity equations (\ref{NS}-\ref{continuity}) to lubrication equations and boundary conditions. We briefly verbally summarize the standard aspects (which can be in found in \cite{leal2007advanced}, for example), and provide detail on what is new due to substrate curvature. The lubrication approximation retains a balance in the tangential momentum equation between the pressure term and the largest viscous term, involving two normal derivatives of tangential velocity. This gives rise to the pressure scaling already introduced above in Section \ref{sec:scalings}. Dropped terms include those multiplied by a Reynolds number, which we assume is small enough not to affect the scalings. The characteristic velocity $u_c$ will be at least as small as $O(\varepsilon^2)$, but may be smaller.  The terms that we will retain are lowest order, including anything that \emph{might} be lowest order once the characteristic velocity is chosen to reflect the governing physics--- the leading terms corresponding to each type of force. The leading order equations for momentum and continuity, with the former resolved onto tangential and normal directions, are \begin{align}
	{\varepsilon}^2 \frac{\rho G l_c^2}{\mu u_c} \bm{G}_\sT - \nabla_\sS p +  \partial_\sN^2 \bm{u}_\sT = O(\varepsilon, \varepsilon \tfrac{l_c}{R_c}) \, , \label{NS2tangential} \\
	{\varepsilon}^3 \frac{\rho G l_c^2}{\mu u_c} G_\sN - \partial_\sN p = O(\varepsilon, \varepsilon \tfrac{l_c}{R_c})   \, , \label{NS2normal} \\
	\nabla_\sS \cdot \bm{u}_\sT + \partial_\sN u_\sN = O(\varepsilon, \varepsilon \tfrac{l_c}{R_c}) \, . \label{continuity2}
\end{align}
Again, the order of $\varepsilon$ is an informal and overly conservative reminder that our fields are the lowest order in an asymptotic expansion, while the order of $\varepsilon \tfrac{l_c}{R_c}$ comes from the Taylor truncation of the geometry; this is also overly conservative in Equation \eqref{NS2normal} as directly derived from \eqref{NS}, but we are being consistent across the entire coupled system (\ref{NS2tangential}-\ref{continuity2}). 

The boundary conditions on the substrate are simply that the velocity vanishes there.  With our scalings and conventions, we write
\begin{align}
	\bm{u}_\sT = O(\varepsilon) \quad \text{at} \; N=0 \, , \label{substrateBCt}\\
	u_\sN = O(\varepsilon) \quad \text{at} \; N=0 \, . \label{substrateBCn}
\end{align}
At the free surface, the kinematic condition 
\begin{align}
	\frac{\partial F}{\partial t} = - | \nabla F |\, \bm{u}\cdot\bm{n} \quad \text{at} \; N=h \, 
\end{align}
is approximately 
\begin{align}
	\frac{\partial h}{\partial t} = u_\sN - \bm{u}_\sT\cdot\nabla_\sS h + O(\varepsilon,\varepsilon\tfrac{l_c}{R_c}) \quad \text{at} \; N=h \, ,\label{kinematicBC}
\end{align}
noting however that this entire equation is multiplied by an $\varepsilon$. 
We allow for a surface tension $\sigma$, a function of the tangential coordinates, nondimensionalized by a characteristic value $\sigma_c$.  For simplicity, we assume that on the other side of the boundary is something with negligible viscosity, providing no surface-tangential tractions and only a nondimensional pressure $p^*$. This leads to normal and tangential stress balances
\begin{align}
	p - p^* - \varepsilon^2 \bm{n}\cdot \left[\nabla\bm{u}+(\nabla\bm{u})^\top\right]\cdot\bm{n} = \varepsilon^2 \frac{\sigma_c}{\mu u_c} \sigma \nabla\cdot\bm{n} + O(\varepsilon,\varepsilon\tfrac{l_c}{R_c}) \quad \text{at} \; N=h \, , \\
	\varepsilon\, \bm{n}\cdot \left[\nabla\bm{u}+(\nabla\bm{u})^\top\right]\cdot\bm{t}_\alpha = \varepsilon \frac{\sigma_c}{\mu u_c} \nabla\sigma\cdot\bm{t}_\alpha + O(\varepsilon,\varepsilon\tfrac{l_c}{R_c}) \quad \text{at} \; N=h \, ,
\end{align}
which are approximately
\begin{align}
	p - p^* &= -\varepsilon^2 \frac{\sigma_c}{\mu u_c} \, \sigma 
	\left[ 	\tfrac{l_c}{R_c}2H + \varepsilon \tfrac{l_c^2}{R_c^2}\, h (4H^2 - 2K)  + {\varepsilon}\nabla^2_\sS h \right] 
	+ O(\varepsilon,\varepsilon\tfrac{l_c}{R_c}) \quad \text{at} \; N=h \, , \label{normalstressBC} \\
	\partial_\sN\bm{u}_\sT &= \varepsilon \frac{\sigma_c}{\mu u_c} \nabla_\sS\sigma + O(\varepsilon,\varepsilon\tfrac{l_c}{R_c}) \quad \text{at} \; N=h \, . \label{tangentialstressBC}
\end{align}
The normal condition, with terms from the divergence of free surface normals, is an early indication that the scaling of the characteristic velocity $u_c$ may depend on whether the geometry is generic or a special case motivating the retention of higher order terms. The gradient in surface tension in the tangential condition has not yet been rescaled into its final form.  This will happen later, when a surfactant field is introduced in Section \ref{sec:surfactant}.

\section{Evolution equation for the thickness}\label{sec:evolution}

We now use equations (\ref{NS2tangential}-\ref{continuity2}) and boundary conditions 
(\ref{substrateBCt}-\ref{substrateBCn}, \ref{kinematicBC}, \ref{normalstressBC}-\ref{tangentialstressBC})
to generate an evolution equation for the film thickness $h$ that retains only the potentially leading order terms of each type. 
Again we build upon the procedure outlined in Leal \cite{leal2007advanced}, much of this portion of which is standard.  
An important element of this derivation is the direct through-thickness integration of the continuity equation and application of the Leibniz rule. 
We are able to retain the simplicity of this procedure in our low order treatment. 

First, integrate the normal momentum equation \eqref{NS2normal} in $N$ and apply the free surface normal stress boundary condition \eqref{normalstressBC} to obtain the pressure 
\begin{align}
	p - p^* = \varepsilon^3 \frac{\rho Gl_c^2}{\mu u_c}\left[G_\sN(N-h)\right] - \varepsilon^2 \frac{\sigma_c}{\mu u_c} \, \sigma 
	\left[ 	\tfrac{l_c}{R_c}2H + \varepsilon \tfrac{l_c^2}{R_c^2}\, h (4H^2 - 2K)  + {\varepsilon}\nabla^2_\sS h \right] + O(\varepsilon,\varepsilon\tfrac{l_c}{R_c}) \, .
\end{align}
We are going to insert $\nabla_\sS p$ into the tangential momentum equation \eqref{NS2tangential}, integrate it twice in $N$, and apply the free surface tangential stress boundary condition \eqref{tangentialstressBC} and substrate no-slip condition \eqref{substrateBCt} to obtain $\bm{u}_\sS$. Before doing so, we can assess what terms to retain in $\nabla_\sS p$. Our intent is to retain in $\bm{u}_\sS$ only the leading order terms of each type, including those next order terms that may become leading order in particular geometries, with the expectation that $u_c$ will scale so as to make these terms $O(1)$.  The tangential momentum equation \eqref{NS2tangential} and free surface tangential stress boundary condition \eqref{tangentialstressBC} will supply, respectively, terms of the form ${\varepsilon}^2 \frac{\rho G l_c^2}{\mu u_c} \bm{G}_\sT$ and $\varepsilon \frac{\sigma_c}{\mu u_c} \nabla_\sS\sigma$. Two terms of the same type arising in $\nabla_\sS p$ will be of lower order.  To see this, we must observe \cite{thiffeaultkamhawi06} that \(\nabla_\sS G_\sN = \nabla_\sS (\bm{G} \cdot \bm{N}) = - \tfrac{l_c}{R_c} \bm{G}\cdot \bm{b}  = - \tfrac{l_c}{R_c} \bm{G}_\sT\cdot \bm{b} \) is a tangential-gravity term. 
Anticipating our final result, we write (with by now familiar shorthand) 
\begin{align}
	\nabla_\sS p = &-\varepsilon^3 \frac{\rho Gl_c^2}{\mu u_c}\left( G_\sN\nabla_\sS h + O(\tfrac{l_c}{R_c})\bm{G}_\sT \right)  \nonumber \\
	&- \varepsilon^2 \frac{\sigma_c}{\mu u_c} \, \sigma \nabla_\sS
	\left[ 	\tfrac{l_c}{R_c}2H + \varepsilon \tfrac{l_c^2}{R_c^2}\, h (4H^2 - 2K)  + \varepsilon\nabla^2_\sS h \right] \nonumber \\
	&+O(\varepsilon^2\tfrac{l_c}{R_c},\varepsilon^3)\frac{\sigma_c}{\mu u_c} \nabla_\sS\sigma + O(\varepsilon,\varepsilon\tfrac{l_c}{R_c}) \, .\label{gradp}
\end{align}
 The tangential gravity term will be dropped, as it is higher order than the corresponding term to be supplied by the tangential momentum equation. The normal gravity term will be retained, as it is the relevant driving term for a planar surface oriented perpendicularly to gravity, such that the tangential gravity term disappears. Two higher order capillary terms from the Laplacian of free surface normals are retained because they will be leading order should $\nabla_\sS H$ vanish, as it does on constant mean curvature surfaces. In the middle coupling term, we choose to keep the trace of the square of the curvature tensor $b_\alpha^\beta b^\alpha_\beta = 4H^2-2K$ intact for the sake of clarity, although dropping pieces of it could also be justified. The term involving the gradient of surface tension will be dropped, as it is higher order than the corresponding term to be supplied by the free surface tangential stress condition. 
The two dropped terms are such that we do not anticipate any physical or geometrical scenario in which $u_c$ will scale so as to make these terms $O(1)$, as there will always be a term of the same type at a lower order of $\varepsilon$ that can set such a scale for $u_c$.  
Therefore, these terms will not be retained further. Finally, insert, integrate, and apply boundary conditions as indicated above to obtain
\begin{align}
	\bm{u}_\sT = &\;\varepsilon^2\frac{\rho G l_c^2}{\mu u_c}  (Nh - \tfrac{N^2}{2}) \left(  \bm{G}_\sT + \varepsilon G_\sN \nabla_\sS h \right)  \nonumber \\
&+\varepsilon^2 \frac{ \sigma_c}{\mu u_c}  ( Nh - \tfrac{N^2}{2} )\, \sigma \nabla_\sS
	\left[ 	\tfrac{l_c}{R_c}2H + \varepsilon \tfrac{l_c^2}{R_c^2}\, h (4H^2 - 2K)  + {\varepsilon}\nabla^2_\sS h \right]   \nonumber \\
 &+ \varepsilon\frac{\sigma_c}{\mu u_c} N \nabla_\sS \sigma + O(\varepsilon, \varepsilon\tfrac{l_c}{R_c}) \, . \label{us}
\end{align}
Recall that although this velocity field is a function of $N$, the gradients in the expression \eqref{us} use the substrate basis, so normal integrations will see only the polynomial coefficients in $N$. 
These approximations allow the application of Leal's straightforward process for planar substrates \cite{leal2007advanced}. 
 Integrate the continuity equation \eqref{continuity2} in $N$ from $0$ to $h$, apply the boundary conditions \eqref{substrateBCn} and \eqref{kinematicBC} on the normal velocity $u_\sN$, and employ the Leibniz rule $\nabla_\sS\cdot\int_0^h\bm{u}_\sT\, \mathrm{d}N = \int_0^h\nabla_\sS\cdot\bm{u}_\sT\, \mathrm{d}N + \bm{u}_\sT |_{N=h}\cdot \nabla_\sS h$  to obtain
\begin{align}
	 \frac{\partial h}{\partial t} = - \nabla_\sS \cdot \int_0^{h(\eta_1, \eta_2, t)} \bm{u}_\sT\, \text{d}N + {O({\varepsilon}, \varepsilon \tfrac{l_c}{R_c})} \, . \label{leibniz}
\end{align}
Finally, insert \eqref{us} and integrate: 
\begin{align}
	\frac{\partial h}{\partial t} = - \nabla_\sS \cdot \Big( &\varepsilon^2\frac{\rho G l_c^2}{\mu u_c} \frac{h^3}{3}\left[ \bm{G}_\sT + \varepsilon G_\sN \nabla_\sS h \right] \nonumber \\
&+\varepsilon^2 \frac{ \sigma_c}{\mu u_c}  \frac{h^3}{3} \sigma \nabla_\sS
	\left[ 	\tfrac{l_c}{R_c}2H + \varepsilon \tfrac{l_c^2}{R_c^2}\, h (4H^2 - 2K)  + {\varepsilon}\nabla^2_\sS h \right] \nonumber \\
&+ \varepsilon\frac{\sigma_c}{\mu u_c} \frac{h^2}{2}\nabla_\sS \sigma \Big) 
+ {O({\varepsilon}, \varepsilon \tfrac{l_c}{R_c})} \, . \label{dhdt}
 \end{align}
Just as we keep the capillary coupling term intact, we keep the normal gravity component $G_\sN$ where it is for clarity of expression, although should it become important due to vanishing of tangential gravity, its derivative will also vanish and it could be moved outside the outer divergence. 

The expression \eqref{dhdt} is minimal, in that it contains no more than is necessary to treat the leading order effects of the three types of forces we are considering, for any substrate geometry and orientation.  With some difficulty, one may compare with \cite{RoySchwartz97, thiffeaultkamhawi06} who include many higher order terms.  In addition to the terms dropped from the pressure gradient \eqref{gradp} due to their relative size, some higher order terms were avoided by direct integration of the leading order continuity equation and application of a one-dimensional Leibniz rule. The presence, as found in many derivations, of nonlinear terms in $h$ on the left hand side, and related terms on the right hand side, is due to complications associated with the flux through a curvilinear volume over the substrate. While such terms can be subsequently dropped by inspection, as pointed out for example in \cite{Howell03, Kang17}, it is preferable not to have them to begin with.  We note that such simplicity is also found in Myers, Charpin, and Chapman \cite{Myers02}, and also that others \cite{SchwartzWeidner95, TakagiHuppert10, Balestra16} have directly invoked a simpler approximate mass conservation relation for one-dimensional or axisymmetric cases that bypasses these issues.

\section{Coupling to a dilute, non-diffusing surfactant}\label{sec:surfactant}

As a simple example of the way that gradients in surface tension might couple to another dynamic field, we consider a surfactant concentration $\Gamma$, a function of the tangential coordinates, nondimensionalized by a characteristic $\Gamma_c$, and the (dimensionful) linear equation of state 
\mbox{$\sigma_c\sigma = \sigma_0 - \Phi\Gamma_c\Gamma = \sigma_m - \Phi\Gamma_c(\Gamma - \Gamma_m)$}, 
where $\sigma_0$ is the value of the surface tension in the absence of surfactant, $\sigma_m$ is a mean value associated with a mean value of surfactant $\Gamma_m$, and $\Phi$ is a positive coefficient. 
Associating $\sigma_c$ and $\Gamma_c$ with these mean values leads to the dimensionless form
\begin{align}
	\sigma = 1 - \Phi\frac{\Gamma_m}{\sigma_m}(\Gamma - \Gamma_m) \, .
\end{align}
Following common precedent, we are going to consider this to be a small perturbation around the mean value of $\sigma$, so that the $\sigma_c\sigma$ in the capillary terms will be approximated by a constant $\sigma_m$ incorporated into the nondimensional coefficient, while the gradient of $\sigma$ will become a negative gradient of $\Gamma$ associated with another nondimensional coefficient.  

Neglecting any diffusion 
or bulk solubility of surfactant, its concentration evolves according to 
\begin{align}
	\frac{\partial \Gamma}{\partial t}  &= -\bm{u}_\sT\cdot\nabla \Gamma - \Gamma(\bm{t}_\alpha\bm{t}^\alpha\cdot\nabla)\cdot\bm{u}  \nonumber \\ 
	&=  -\bm{u}_\sT\cdot\nabla \Gamma - \Gamma(\bm{t}_\alpha\bm{t}^\alpha\cdot\nabla)\cdot( \bm{t}_\alpha\bm{t}^\alpha\cdot\bm{u} ) - \Gamma({\nabla} \cdot \bm{n})(\bm{n}\cdot\bm{u}) \, , \label{surftrans} 
\end{align}
where $\nabla$ and $\bm{u}$ are evaluated at $N=h$, and the partial time derivative is the same rate of change in the fixed substrate-adapted coordinates used in all the other equations.  Note that $\nabla \Gamma = \bm{t}_\alpha\bm{t}^\alpha\cdot\nabla\Gamma$. 
The typical and more natural way to express this transport equation instead uses a subscripted $S$ to refer to the free surface, and a partial time derivative that follows its normal motion, so that the total time derivative of $\Gamma$ is dissected differently and the right hand side can be combined into a single free surface divergence. 
The approximation of equation \eqref{surftrans} loses the last term and uses a single substrate divergence, 
\begin{align}
	\frac{\partial \Gamma}{\partial t} = - \nabla_\sS \cdot \Gamma \bm{u}_\sT  |_{N=h} + O(\varepsilon, \varepsilon \tfrac{l_c}{R_c}) \, ,\label{surftrans2}
\end{align}
and this will accompany the thickness equation on the substrate by expressing $\bm{u}_\sT |_{N=h}$ as \mbox{$\bm{u} |_{N=h}\cdot\bm{a}_\alpha\bm{a}^\alpha$}. Such a pair of coupled equations is used in Li and Kumar \cite{LiKumar18}, who also include diffusion.
In Problem 6--12(a) of Leal \cite{leal2007advanced}, the student is asked to derive a similar surfactant equation for a planar substrate. 
Some of the velocity field in \eqref{surftrans2} will arise from Marangoni forces due to gradients in surfactant concentration, leading to nonlinearities in $\Gamma$.  Like others before us, for example \cite{LiKumar18}, we do not attempt to break $\Gamma$ into its mean and variable parts, as they may be of the same order of magnitude. The small quantity here is $\Phi\frac{\Gamma_m}{\sigma_m}$, which also implies that either $\sigma_0$ or $\sigma_m$ could serve as $\sigma_c$. If instead one makes the assumption that $\Gamma/\Gamma_m$ is close to unity, this would imply stiff behavior, as for a nearly uniform surfactant concentration preserved by a velocity field on the interface that is incompressible at leading order.

\section{Nondimensional groups and general equations}\label{sec:groups}

The nondimensional groups that arise in this derivation are: the gravity number $\textsf{G}\equiv\frac{\rho G\l_c^2}{\mu u_c}$ between gravitational and viscous forces (see \cite[expression 2.37c]{Oron97}, but this is often unnamed in the literature, and should of course not be confused with the magnitude of gravitational acceleration $G$), 
the inverse capillary number $\textsf{Ca}^{-1}\equiv \frac{\sigma_m}{\mu u_c}$ comparing surface tension and viscous forces, and the Marangoni number $\textsf{Ma}\equiv\frac{\Phi\Gamma_m}{\mu u_c}$ comparing surface tension gradient and viscous forces (this is one of many common definitions, see \cite[Table 3]{ManikantanSquires20}).  
Also relevant are the ratios of these quantities: the Bond number $\textsf{Bo}=\textsf{GCa}$ or ratio of gravitational and capillary forces, an alternate definition of a Marangoni number $\textsf{Ma}_\sigma\equiv\textsf{CaMa}$ or ratio of capillary and Marangoni forces (surface tension and surface tension gradient forces, see again \cite[Table 3]{ManikantanSquires20}), and finally a number $\textsf{Ma}^{-1}\textsf{G}$ (=$\textsf{Ma}_\sigma^{-1}\textsf{Bo}$) that compares Marangoni and gravitational forces, a quantity likely relevant to ``wine legs'' 
(see the unnamed dimensionless group in \cite[Table 4]{HosoiBush01}). 

In terms of these quantities, the coupled equations may be written 
\begin{align}
	\frac{\partial h}{\partial t} = - \nabla_\sS \cdot \Big( &\varepsilon^2\textsf{G} \frac{h^3}{3}\left[ \bm{G}_\sT + \varepsilon G_\sN \nabla_\sS h \right] \nonumber \\
		&+\varepsilon^2 \textsf{Ca}^{-1} \frac{h^3}{3} \nabla_\sS
	\left[ 	\tfrac{l_c}{R_c}2H + \varepsilon \tfrac{l_c^2}{R_c^2}\, h (4H^2 - 2K)  + {\varepsilon}\nabla^2_\sS h \right] \nonumber \\
		&- \varepsilon \textsf{Ma} \frac{h^2}{2}\nabla_\sS \Gamma \Big) 
		+ {O({\varepsilon}, \varepsilon \tfrac{l_c}{R_c})} \, , \label{dhdtnon}\\
	\frac{\partial \Gamma}{\partial t} = - \nabla_\sS \cdot \Gamma &\bm{u}_\sT  |_{N=h} + O(\varepsilon, \varepsilon \tfrac{l_c}{R_c}) \, , \label{surftransnon}\\
	\bm{u}_\sT |_{N=h} = 
	&\;\varepsilon^2\textsf{G} \frac{h^2}{2}\left[ \bm{G}_\sT + \varepsilon G_\sN \nabla_\sS h \right] \nonumber \\
		&+\varepsilon^2 \textsf{Ca}^{-1} \frac{h^2}{2} \sigma \nabla_\sS
	\left[ 	\tfrac{l_c}{R_c}2H + \varepsilon \tfrac{l_c^2}{R_c^2}\, h (4H^2 - 2K)  + \varepsilon\nabla^2_\sS h \right] \nonumber \\
		&- \varepsilon \textsf{Ma} h \nabla_\sS \Gamma + O(\varepsilon, \varepsilon \tfrac{l_c}{R_c}) \, . \label{usnh} 
\end{align}
The nondimensional numbers all contain a factor of $u_c$ in their denominators.  Generically we expect $u_c$ to carry an $\varepsilon^2$ so that one or more of $\varepsilon^2\textsf{G}$ or $\varepsilon^2\textsf{Ca}^{-1}$ will be $O(1)$.  Special cases with respect to orientation (gravity) or shape (capillarity) would imply $u_c$ to carry an $\varepsilon^3$.  For Marangoni-driven thin film flows, we might still expect $u_c$ to carry an $\varepsilon^2$,  
while $\Phi\Gamma_m$ would provide an $\varepsilon$, so that $\varepsilon\textsf{Ma}$ will be $O(1)$. 
The absorption of the constant mean $\sigma_m$ into the inverse capillary number is in keeping with \cite{Kang17, LiKumar18}, and some dropped terms related to nonconstant $\sigma$ can now be recognized as $O(\varepsilon^2\tfrac{l_c}{R_c}\textsf{Ma},\varepsilon^3\textsf{Ma})$.  

The system (\ref{dhdtnon}-\ref{usnh}) will serve as a reference for specific examples.  It is not intended to be used in its entirety.  If the geometry is generic, only the leading order gravity and capillary terms, namely the tangential gravity and substrate mean curvature gradient terms, should be retained.  An example of an exceptional case is a planar substrate oriented perpendicularly to gravity, such that both tangential gravity and mean curvature disappear, and thus the normal gravity term and the Laplacian of free surface thickness are important. Other possibilities and further details will be found below.

\section{Examples of specific forms of the equations}\label{sec:examples}

At risk of belaboring the point, we present several possible employable forms of the equation system (\ref{dhdtnon}-\ref{usnh}) applied to  settings of potential interest. This is far from an exhaustive list. 
From here forward, we will drop the appended $O(\varepsilon, \varepsilon \tfrac{l_c}{R_c})$ on each of the expressions, although it should be remembered that this approximation still holds in each case.

\subsection{Generic substrates}\label{sec:generic}

A generic surface is one such that $\nabla_\sS H$ is $O(1)$, which is the case for all but some of the most common and simple surfaces. If all three types of driving forces are relevant, we have the coupled system of two equations:  
\begin{align}
	\frac{\partial h}{\partial t} &= - \nabla_\sS \cdot \Big( \varepsilon^2\textsf{G} \frac{h^3}{3}  \bm{G}_\sT 
	+\varepsilon^2 \tfrac{l_c}{R_c}\textsf{Ca}^{-1}  \frac{h^3}{3} \nabla_\sS 2H - \varepsilon \textsf{Ma} \frac{h^2}{2}\nabla_\sS \Gamma \Big) \, , \label{dhdtgeneric}\\
	\frac{\partial \Gamma}{\partial t} &= - \nabla_\sS \cdot  \Gamma \Big( \varepsilon^2\textsf{G} \frac{h^2}{2}  \bm{G}_\sT 
	+\varepsilon^2 \tfrac{l_c}{R_c} \textsf{Ca}^{-1} \frac{h^2}{2} \nabla_\sS 2H - \varepsilon \textsf{Ma} h\nabla_\sS \Gamma  \Big) \, .\label{surftransgeneric}
\end{align}
These include the leading order effects of gravity, capillarity (surface tension), and Marangoni (surface tension gradient) forces. Clearly, if Marangoni effects are not present, the system reduces to the single equation
\begin{align}
	\frac{\partial h}{\partial t} = - \nabla_\sS \cdot \Big( \varepsilon^2\textsf{G} \frac{h^3}{3}  \bm{G}_\sT 
	+\varepsilon^2 \textsf{Ca}^{-1} \tfrac{l_c}{R_c} \frac{h^3}{3} \nabla_\sS 2H \Big) \, ,  \label{dhdtgenericnoMa}
\end{align}
where the surface tension is now merely a constant inside the capillary number. 
Recall also that $\nabla_\sS\cdot\bm{G}_\sT  =  \tfrac{l_c}{R_c} 2HG_\sN$.  
 The capillary term is what Howell \cite{Howell03} calls the large curvature case. His small curvature case also retains the free surface curvature term $\nabla^2_\sS h$, assumed to be of the same order as the substrate curvature. Both of these cases and their significance are also addressed by Myers, Charpin, and Chapman \cite{Myers02}. 
 
Howell's small curvature case is invoked by \cite{Lin21} in an analysis of a conical funnel, including gravity. This small curvature treatment should only be valid for shallow funnels, which would also justify their inclusion of normal gravity terms at the same order as tangential. In contrast, \cite{XueStone21} in their treatment of a conical funnel recognize that the normal gravity term is small, but miss a capillary curvature coupling term, which is only valid if the substrate curvature is assumed small. This is explicitly stated in \cite{Myers02} in a critique of   \cite{SchwartzWeidner95}, who miss this coupling term in their derivation for general one-dimensional surfaces.  Further detail may be found in Sections \ref{sec:cmc} and \ref{sec:smallslope} below.

\subsection{Constant mean curvature nonplanar substrates}\label{sec:cmc}

For such cases, the leading order capillary term vanishes and we retain the coupling and free surface terms:  
\begin{align}
	\frac{\partial h}{\partial t} &= - \nabla_\sS \cdot \Big( \varepsilon^2\textsf{G} \frac{h^3}{3}  \bm{G}_\sT
	+\varepsilon^3 \textsf{Ca}^{-1} \frac{h^3}{3} \nabla_\sS\left[  \tfrac{l_c^2}{R_c^2}\, h (4H^2 - 2K)  + \nabla^2_\sS h \right] - \varepsilon \textsf{Ma} \frac{h^2}{2}\nabla_\sS \Gamma \Big) \, ,  \label{dhdtcmc}\\
	\frac{\partial \Gamma}{\partial t} &= - \nabla_\sS \cdot  \Gamma \Big( \varepsilon^2\textsf{G} \frac{h^2}{2}  \bm{G}_\sT 
	+\varepsilon^3 \textsf{Ca}^{-1}  \frac{h^2}{2} \nabla_\sS\left[  \tfrac{l_c^2}{R_c^2}\, h (4H^2 - 2K)  + \nabla^2_\sS h \right] - \varepsilon \textsf{Ma} h\nabla_\sS \Gamma \Big) \, . \label{surftranscmc}
\end{align}
This form is relevant to the equations derived for specific simple shapes in, for example, \cite{Myers02, Kang17, LiKumar18}, or to Howell's more restrictive constant curvature case \cite{Howell03}. 
The trace of the square of the curvature tensor is kept intact here, although pieces of it will vanish. In a lines of curvature derivation, this term is expressed as the sum of the squares of the principal curvatures \cite{Myers02, Howell03}. 
For a circular cylinder of radius $R$, $4H^2-2K = 1/R^2$.  
For a sphere of radius $R$, $4H^2-2K = 2/R^2$. 
However, in general, constant $H$ surfaces have nonconstant $K$, and the form of the equations does not considerably simplify. 
These surfaces have the interesting property that it does not matter at leading order whether the flows are above or below, inside or outside, the substrate. 
Thus there exists a symmetry between interior (rimming) and exterior (coating) flows on a cylinder \cite{PreziosiJoseph88}. 
This is not true for a perpendicular case for which the normal gravity term is retained, or a generic surface for which the dominant capillary term is the gradient of mean curvature. One might also imagine a small deviation from a constant mean curvature nonplanar substrate, where all three capillary terms might be important. Relevant discussion may also be found in \cite{Myers02, Howell03}.

\subsection{Planar substrates}\label{sec:planar}

For a planar substrate, $\tfrac{l_c}{R_c}=0$.  Terms that might remain include: 
\begin{align}
	\frac{\partial h}{\partial t} &= - \nabla_\sS \cdot \Big( \varepsilon^2\textsf{G} \frac{h^3}{3}\left[ \bm{G}_\sT + \varepsilon G_\sN \nabla_\sS h \right] + \varepsilon^3 \textsf{Ca}^{-1} \frac{h^3}{3} \nabla_\sS^3 h - \varepsilon \textsf{Ma} \frac{h^2}{2}\nabla_\sS \Gamma \Big)  \, , \label{dhdtplanartilt}\\
	\frac{\partial \Gamma}{\partial t} &= 
	 - \nabla_\sS \cdot  \Gamma \Big( \varepsilon^2\textsf{G} \frac{h^2}{2}\left[ \bm{G}_\sT + \varepsilon G_\sN \nabla_\sS h \right]  
	+\varepsilon^3 \textsf{Ca}^{-1}  \frac{h^2}{2} \nabla_\sS^3 h - \varepsilon \textsf{Ma} h\nabla_\sS \Gamma \Big) \, . \label{surftransplanartilt}
\end{align}
where $\nabla_\sS^3 \equiv \nabla_\sS \nabla_\sS \cdot \nabla_\sS$. 
The evaluation of these expressions will simplify further, as $\nabla_\sS$ is Cartesian, and $\bm{G}_\sT$ and $G_\sN$ can be moved out of the divergences, as they are constants, trigonometric functions of a tilt angle. 
There is an extensive literature on such problems, often in a one-dimensional setting, which we do not address in this paper. Depending on the tilt angle and Bond number, certain terms may be neglected. For generic tilt angles, the normal gravity term may be neglected, and inclusion of the capillary term would imply $\textsf{Bo}$ to be $O(\varepsilon)$. For small tilt angles, tangential gravity $\bm{G}_\sT$ may become $O(\varepsilon)$; for zero tilt, it vanishes.

\subsection{Small slope, nearly planar substrates}\label{sec:smallslope}

For a nearly planar, but wavy surface, $\tfrac{l_c}{R_c}$ is small. 
This is the small curvature case \cite{Myers02, Howell03} mentioned above in Section \ref{sec:generic}. 
The coupling term is neglected, and we retain the Laplacian terms of the substrate (the mean curvature) and free surface.  
All the other terms might remain: 
\begin{align}
	\frac{\partial h}{\partial t} &= - \nabla_\sS \cdot \Big( \varepsilon^2\textsf{G} \frac{h^3}{3}\left[ \bm{G}_\sT + \varepsilon G_\sN \nabla_\sS h \right] +\varepsilon^2 \textsf{Ca}^{-1} \frac{h^3}{3} \nabla_\sS\left[  \tfrac{l_c}{R_c}\,2H + \varepsilon\nabla^2_\sS h \right]  - \varepsilon \textsf{Ma} \frac{h^2}{2}\nabla_\sS \Gamma \Big) \, .  \label{dhdtss}\\
	\frac{\partial \Gamma}{\partial t} &= 
	 - \nabla_\sS \cdot  \Gamma \Big( \varepsilon^2\textsf{G} \frac{h^2}{2}\left[ \bm{G}_\sT + \varepsilon G_\sN \nabla_\sS h \right]  
	+\varepsilon^2 \textsf{Ca}^{-1}  \frac{h^2}{2} \nabla_\sS\left[ \tfrac{l_c}{R_c}\,2H + \varepsilon\nabla^2_\sS h \right]  - \varepsilon \textsf{Ma} h\nabla_\sS \Gamma \Big) \, . \label{surftransss}
\end{align}
Further simplification might entail breaking $\bm{G}_\sT$ and $G_\sN$ into constants plus a correction; their derivatives carry an additional order of $O(\tfrac{l_c}{R_c})$. 
One might further assume that the constant part of $\bm{G}_\sT$ is zero, so that the substrate is both nearly planar and nearly perpendicular to gravity.  
If both $\tfrac{l_c}{R_c}$ and the constant part of $\bm{G}_\sT$ are $O(\varepsilon)$, we might retain all the terms in the above equation. 
These types of simplifications are implicit in many works, with which one may compare with some difficulty, making note of possible differences and keeping in mind that $G_\sN$ is typically negative.  Treatments of spin coating have an additional tangential driving term \cite{HwangMa89, StillwagonLarson90, Kalliadasis00, Gramlich02}. 
Some treatments of flows down bumpy hills \cite{Gaskell04, Hinton19} use a tilted Cartesian frame and end up with an additional term for substrate height, because free surface height is measured normal to a tilted plane rather than to the substrate (film thickness). This is effectively like breaking $\bm{G}_\sT$ into constant and nonconstant parts.

\section{Concluding remarks}\label{sec:discussion}

The present derivation, which also serves to collect and compare several known results, is intended to provide a basis for further exploration of the role of substrate geometry in modifying and mediating balances of driving forces in thin films with one free surface. The general equations (\ref{dhdtnon}-\ref{usnh}) retain a minimal set of terms from several sources--- those of leading order of each type, and only those at higher order that may become leading order for special orientations and geometries. They are simple enough to be of potential use in analytical work. 
The system of two coupled conservation laws (\ref{dhdtgeneric}-\ref{surftransgeneric}) is interesting in its own right, and may stimulate further study. 

However, our treatment is missing any consideration of inertia, as typically arises from the rotation of the substrate.  Recent works \cite{Duruk23, Shepherd24} indicate how such an effect may be included in a general geometric setting. Our treatment of Marangoni effects was kept as basic as possible in order to provide a simple illustrative example, and clearly more can be done.  Adding a temperature field or diffusion should be straightforward.  Our assumption that changes in surface tension are small due to the diluteness of the surfactant field, rather than small deviations of this field, could be carefully revisited. 

The complications that arise when nongeneric orientations, and particularly geometries, are considered, namely the need to address various special cases with different types of curvature scaling, suggest that additional nondimensional numbers might be useful. 
This is already hinted at by the use of the length scale ratio $l_c/R_c$ here and in Howell \cite{Howell03}. 
Perhaps there is a way to further nondimensionalize aspects of the substrate geometry 
to encapsulate all possibilities, including those not discussed here, in terms of a set of nondimensional numbers. 
This is an atypical way to describe surfaces.  The distinction between mean curvature and its gradient suggests a tempting analogy with capillary and Marangoni numbers. We note that the derivatives of the gravity terms also encode information about the substrate through its full curvature tensor, not just its invariants as in the capillary terms.  The possibility of anisotropic substrate scaling should also be investigated. 
We expect that a more general development with inertial effects may raise similar questions. 

\section*{Acknowledgments}
We thank N. Ribe for comments on an early draft.

\bibliographystyle{unsrt}

\end{document}